# Alternating-Chiral Charge Density Waves and Hybrid Ferrimagnetism in Monolayered NbTe$_2$

Yusong Bai[1#], Guohua Cao[2#], Jinghao Deng[1#], Haomin Fei[2], Xiaoyu Lin[1], Leiqiang Li[2,5], Chao Zhu[1], Zemin Pan[1], Tao Jian[1], Da Huo[1], Zhengbo Cheng[1], Chih-Kang Shih[4], Ping Cui[2,3*], Chendong Zhang[1*], Zhenyu Zhang[2,3]

[1]*School of Physics and Technology, Wuhan University, Wuhan 430072, China*

[2]*International Center for Quantum Design of Functional Materials (ICQD), University of Science and Technology of China, Hefei, Anhui 230026, China*

[3]*Hefei National Laboratory, University of Science and Technology of China, Hefei, Anhui 230088, China*

[4] *The University of Texas at Austin, Austin TX 78712, USA*

[5] *MinJiang Collaborative Center for Theoretical Physics, College of Physics and Electronic Information Engineering, Minjiang University, Fuzhou 350108, China*

*Correspondence and requests for materials should be addressed to:
cdzhang@whu.edu.cn (C.D.Z.), cuipg@ustc.edu.cn (P.C.)

**Intertwining of different quantum degrees of freedom manifests exotic quantum phenomena in many-body systems, especially in reduced dimensionality. Here we show that monolayered NbTe$_2$ serves as an ideal platform where lattice, charge, and spin degrees of freedom manifest cooperatively, leading to a new and threading order of chirality. By using spin-polarized scanning tunneling microscopy/spectroscopy, we reveal that the $\sqrt{19} \times \sqrt{19}$ phase of NbTe$_2$ is encoded with both alternating-chiral atomic displacements and charge density waves, characterized by two chiral units of opposite handedness within the reconstructed cell. We show unambiguous evidence for emergent spin polarizations spreading over the primitive cell, with the magnetization orientation synchronized with alternating handedness of chiral order. Our first-principles studies identify the origin of intertwined orders being correlation driven, with the threading order of chirality emerging when the on-site Coulomb repulsion exceeds a critical value. The spin ordering is further shown to be of hybrid ferrimagnetic nature, contributed by the itinerant electrons and localized *d*-orbitals. Collectively, these findings expand the realm of chiral order in correlated electron systems, and facilitate an appealing platform for chiral spintronic and related applications.**

Instabilities near the Fermi surface are inherently driven by many-body interactions in solids, and are accompanied by the emergence of exotic collective orders.[1–5] The intertwining/competition of different macroscopic quantum orders is an important paradigm in condensed matter physics to understand and manipulate emergent collective electron behaviors. For instance, in Cu- or Fe-based materials, particular spin orders (e.g., Mott antiferromagnetism) are believed to be parent states of matter for harboring unconventional superconductivity.[6–8] With dimensionality reduction, such collective orders become even more prominent, especially in layered materials with atomic thickness. One compelling platform is served by the T-phase group-VB transition metal dichalcogenides, e.g., $TaS_2$, $TaSe_2$, and $NbSe_2$, where the synergistic effects of electron-electron and electron-phonon interactions collectively lead to a $\sqrt{13} \times \sqrt{13}$ charge density wave (CDW) order characterized by notable in-plane lattice distortions. Strikingly, within each unit cell, thirteen metal atoms collectively reposition themselves to create the well-known Star of David (SoD) cluster.[9–12] Furthermore, the significant on-site Coulomb repulsion associated with the suppression of inter-unit cell hopping helps to convert these nonmagnetic compounds into correlated magnets,[13,14] with the magnetic moments fairly localized at the SoD centers,[15–17] while the magnetic ground states are still under active debate.[18–20]

In addition to the commonly encountered orderings of lattice, charge and spin, chirality can serve as yet another distinct and intriguing degree of freedom in the formation of long-range order. So far, a homogeneous chiral order on the macro/meso scale length has been frequently observed in materials with a chiral nature. In contrast, alternating chiral order, which refers to the simultaneous existence of both types of chirality with alternation in space, has been less realized in naturally forming materials. The concept of anti- or alternating-chiral order was initially introduced to describe the packing of left- and right-handed helices in polymers or soft condensed matters[21,22] and later actively developed in the applications of artificially designed meta-materials by taking advantages of the chirality-dependent mechanical or acoustical properties.[23–25] In the realm of many-body physics, structural chirality was known to be closely connected to the movements and interactions of spin and charge,[26] stimulating a number of theoretical designs of exotic spin textures based on the anti-chiral order.[27–29] However, alternating-chiral order mediated correlation phenomena have been rather rare in practical experiments. One pioneering example was reported in $URu_2Si_2$,[30] where alternating left- and right-handed *f* orbitals at the uranium sites were attributed to a nonmagnetic phase transition at low temperature. The desire to realize structures with different chiral properties is naturally inherited in the emerging field of two-dimensional van der Waals (vdW) materials, as exemplified by the very recent observations of heterochirality via vdW stacking, reflecting the alternation of chirality in two neighboring layers.[31,32]

In this work, combining scanning tunneling microscopy/spectroscopy (STM/S) and density functional theory (DFT) calculations, we reveal that in the recently discovered $\sqrt{19} \times \sqrt{19}$ CDW structure of monolayered (ML) $NbTe_2$, both mirror and inversion symmetries break, thus giving rise to a chiral nature to the CDW order.[33,34] More strikingly, owing to the presence of non-radial displacements, a pair of structural chiral sites with opposite handedness, *i.e.* counterclockwise and clockwise, appears within a reconstructed primitive cell. As a result, an in-plane alternating chiral order with triangular checkerboard patterns emerges, dubbed alterchirality. Moreover, in contrast to the local magnetic moments in the SoD CDW phase,[15–17] our spin-polarized STM/S measurements ascertain the presence of finite magnetic moments spreading over the $\sqrt{19} \times \sqrt{19}$ superstructure, and the chirality order strictly correlates with the magnetization orientation, endowing the system with alternating magnetism in nature as well. Our first-principles calculations further help to identify the physical origin of the above experimental observations to be vitally correlation driven, with the threading order of chirality emerging only when the on-site Coulomb repulsion exceeds a critical value. The spin ordering is further shown to be of a hybrid ferrimagnetic nature, contributed by the itinerant electrons and localized *d*-orbitals. Collectively, these findings expand the realm of chiral order in correlated electron systems, and provide appealing platforms for chiral spintronic[26,35] and related applications.

Among the family of group-VB dichalcogenides, the collective electronic behaviors of NbTe$_2$ have been less explored experimentally in the monolayer limit, while its bulk form is known to be a 3 × 1 CDW superconductor with $T_C$ = 0.7 K.[35,36] Recent studies have shown that the $\sqrt{19} \times \sqrt{19}$ CDW ordered phase is the most stable structure of ML NbTe$_2$ at sufficiently high temperatures.[37,38] As shown in supplementary Fig. S1, here we further find that the $\sqrt{19} \times \sqrt{19}$ phase of NbTe$_2$ can only exist in the monolayer limit, while an earlier study has shown that it can be stabilized in few-layer TaTe$_2$.[39] In principle, the $\sqrt{19} \times \sqrt{19}$ super-periodicity can be considered as an expanded SoD, as collectively described by the formula that the total number of metal atoms per unit cell is $6n + 1$,[40] where $n$ is the number of shells surrounding the central atom (*i.e.*, $n = 2$ for the $\sqrt{13} \times \sqrt{13}$ CDW, and $n = 3$ for the $\sqrt{19} \times \sqrt{19}$ CDW; see Fig. 1a).

Figure 1b shows a large-scale STM image of a 1*T*-NbTe$_2$ monolayer grown on a bilayer graphene/SiC(0001) substrate (see growth details in Methods), demonstrating controlled growth of macroscopic samples with a uniform $\sqrt{19} \times \sqrt{19}$ phase. Within a wide range of negative sample bias, this superstructure is visualized as being composed of three representative spot sites distinguished by their relative topographic heights (see, for example, Fig. 1c at -1.0 V). The bias-dependent morphological evolution for these three spots is illustrated in supplementary Fig. S2. A rhombohedral unit cell of the $\sqrt{19} \times \sqrt{19}$ superstructure is labeled in Fig. 1c with the corners located on the brightest spots. The primitive vector is measured to be ~15.8 Å, while the atomic lattice constant in a pristine NbTe$_2$ cell is measured to be 3.64 Å.[37] In Fig. 1d, the Fourier fast transformed (FFT) pattern acquired from a high-resolution image (inset in Fig. 1d) explicitly manifests that the rotation angle between the basis vectors of the CDW pattern and atomic lattice is 23.4°, consistent with the schematic model in Fig. 1a. Furthermore, we performed Fourier transformation analysis on a series of bias-dependent *dI/dV* maps (see supplementary Fig. S3) and normalized the intensity of CDW peaks ($I_{CDW}$) with respect to the intensity of $1 \times 1$ Bragg peaks ($I_{1 \times 1}$). As shown in Fig. 1e, the ratio $R = \log(I_{CDW}/I_{1 \times 1})$ as a function of the energy well resembles the general tendency in the *dI/dV* spectrum, where $R$ has large amplitudes at energies far from the Fermi level ($E_F$), but shows a two orders of magnitude reduction near $E_F$. These results confirm that the spectroscopic dip corresponds to the CDW gap with a magnitude of $2\Delta_{CDW} \approx 100$ meV.[41,42] Following the mean field formula,[1] $2\Delta = \alpha k_B T$, the transition temperature of the $\sqrt{19} \times \sqrt{19}$ CDW is estimated to be ~331 K in the weak coupling limit ($\alpha = 3.5$), in agreement with the feature shown in our STM image obtained also at room temperature (supplementary Fig. S4).

Figure 1f shows the calculated phonon dispersion of the pristine 1$T$-NbTe$_2$ monolayer, from which we could find several soft modes, with the lowest lying one appearing near the wave vector 0.25$b$ ($b = 2\pi/a^*(1,0)$) along the Γ-M direction. The existence of such a soft mode implies the system's tendency towards developing local order driven by collective atomic displacements. In addition, the phonon spectrum (supplementary Fig. S5) shows that although the reciprocal vectors of the $\sqrt{19} \times \sqrt{19}$ superlattice appear to be slightly misaligned with that of the minimum frequency mode, these vectors are still deeply located within the unstable region, and are therefore expected to have a significant impact on the lattice dynamics. Based on these observations, electron-phonon coupling should play an important role in the formation of CDW orders, likely jointly with the Fermi surface nesting effects.[38]

Figure 2a displays a magnified image with atomic resolution taken at a close tip-to-sample separation ($V_{bias}$ = +10 mV, $I_t$ = 5 nA; and a similar image in large scale is seen in supplementary Fig. S2a). The relaxed $\sqrt{19} \times \sqrt{19}$ atomic structure obtained from DFT calculations is displayed side by side and partially superimposed with the experimental one. A cluster containing 19 Nb atoms can be outlined by the blue dashed lines, exhibiting an apparent deviation from an ideal hexagon with six-fold symmetry. Based on the close agreement between the experimental and theoretical results of bias-dependent topographic contrasts (supplementary Fig. S2b-g), the supercell defined in Fig. 1c can be drawn on the atomic model as labeled. The corner of the rhombohedral cell (brightest spot in Fig. 1c) corresponds to the center of the 19-Nb cluster, referred to as the "center" site. The two hollow sites can be identified as one bright hollow ("BH", centered on a bottom-layer Te atom) and one dark hollow ("DH", centered on a top-layer Te atom). It is intriguing that the DH site manifests as a six-bladed propeller with counterclockwise rotation (Fig. 2a and supplementary Fig. S2a).

To quantify the experimentally observed atomic distortions, the 19 top-layer Te atoms within a CDW supercell are indexed as Te #1-#19. The atoms from Te #1 to #12 are located inside a 19-Nb-atom cluster, forming a truncated hexagon with three long and three short edges. Our measurements reveal that the distances between the three inner Te atoms (e.g., Te #1, #2, and #3) are uniformly compressed to 3.0 Å, notably shorter than ~ 3.64 Å in pristine 1$T$-NbTe$_2$, suggesting a radial contraction towards the central Nb atom. The situation for the outer-shell Te atoms (e.g., from Te #4 to #12) is different. As seen from the line profiles in Fig. 2b, the distances between the Te atoms along the truncated hexagon edges exhibit substantial variations. The short edges (marked by the cyan lines) are significantly stretched with an average Te-Te distance of 4.20 Å, while the long edges (marked by the pink lines) are compressed with an average Te-Te distance of 3.25 Å. The statistical distributions of Te-Te distances over a large-scale image (containing more than 50 supercells) are shown in Fig. 2c and 2d. These experimental values are in reasonably good agreement with the calculated ones where the average Te-Te distances along the stretched and compressed edges are 3.91 and 3.51 Å, respectively. The calculated interatomic distance among Te #1-#3 is 3.48 Å, also shortened from the pristine value, as qualitatively observed experimentally. Our quantitative analyses for other Te atoms near the BH site (e.g., from Te #13 to #18) can be found in supplementary Fig. S6, which further demonstrate the overall consistency between the experimental and theoretical results.

In the $\sqrt{19} \times \sqrt{19}$ CDW phase of the NbTe$_2$ monolayer, the divergent stretched *vs.* contracted atomic displacements along the originally equivalent zig-zag directions suggest non-radial distortions (Fig. 2a), contrary to the universal radial contraction in the $\sqrt{13} \times \sqrt{13}$ CDW phase.[9-11] As shown in Fig. 2e, the arrows represent the in-plane displacement vectors extracted by comparing the $\sqrt{19} \times \sqrt{19}$ superstructure with the pristine 1*T*-NbTe$_2$ structure. Distinctive behaviors can be visibly revealed for the center, BH, and DH sites. Within each region, the displacements of Te and Nb atoms are concerted and exhibit qualitatively similar tendencies, and for clarity in our subsequent presentations only the Nb atoms are depicted. In the center region, seven Nb atoms are enclosed in a green shadow, and the radial contraction towards the fixed central Nb atoms is the dominant movement. More strikingly, we find that the rotations of the vectors at the DH (masked with the cyan shadow) and BH regions (masked with the pink shadow) form a pair of alternating chiral vortices, corresponding to counterclockwise and clockwise vortices, respectively. Therefore, the observed CDW phase possesses a 2D alternating chiral order with the local chirality centered on the two triangular half supercells associated with alternating handedness. In analogy to the chiral CDW phase of the √13 × √13 SoD,[33,34] here the novel observed chiral √19 × √19 superstructure can be classified as an alterchiral CDW.

To gain some insight into the microscopic driving force for the alterchiral CDW, we systematically investigate the dependence of the structure on various physical factors, including the effects of strain, interfacial coupling, and electron correlations using DFT calculations. We find that, under comparable physical conditions, the on-site Coulomb repulsion plays a vital role in inducing the CDW superstructure. As shown in supplementary Fig. S7a, when the on-site Coulomb repulsion $U = 0$, the Te-Te distances along the stretched (3.68 Å) and compressed (3.55 Å) edges are very close, and the superstructure is barely chiral. In contrast, the distinct alterchiral superstructure emerges from the √19 × √19 CDW phase only when $U$ is large enough (see supplementary Fig. S7b). Consequently, the underlying physical origin of the alterchirality is rooted in electron correlations.

We next move to the investigation of magnetic properties of the √19 × √19 CDW phase by using spin-polarized Cr-coated W tips. Thick Cr coating (40 nm in our works) tends to have in-plane magnetization (namely, perpendicular to the tip axis), which is resistant against the external magnetic field, and yields a relatively small stray field.[43] In principle, the spin-polarized signal is proportional to $M_T M_S cos\theta$, where $M_T$ and $M_S$ are the magnetic moments in the tip and sample, respectively, and $\theta$ is the angle between $M_T$ and $M_S$ and can be tuned by an external field.[43,44] Figure 3a-c plots the normalized $dI/dV$ spectra at the center, BH, and DH sites under different perpendicular magnetic fields of $B$ = -2, 0, and +2 T. For all three sites, we obtain the spin-polarized *dI/dV* signals manifested by the differences in the *dI/dV* spectra with and without *B*. Taking the DH site as an example, the *B*-dependent *dI/dV* intensity at the fixed bias of -0.35 V with a maximum polarization is illustrated in Fig. 3d. Here the conductance shows a peak at *B* = 0 T, and is suppressed by applying an external magnetic field. As seen from Fig. 3d, such a suppressing behavior is nearly saturated when $|B| \geq 0.2$ T and is independent of the magnetic field polarity, implying that the CDW phase exhibits in-plane magnetization. These observations can be understood within the following scenario. At *B* = 0 T, both $M_T$ and $M_S$ lie within the basal plane, leading to a conductance peak. While $M_T$ is strongly resistant to the external field, $M_S$ can be easily aligned with the field. Therefore, at a finite $|B|$, the angle $\theta$ equals 90º or -90º, leading to a zero spin-polarized portion in differential conductance. The slight deviation from a symmetric line shape in Fig. 3d is likely to result from the small canting in the tip magnetization, which is common for such Cr-coated tips.

Going beyond the DH site, we show in Fig. 3a-c the polarity of the spin polarization *P* (defined as $2\frac{dI/dV|_{0T} - dI/dV|_{2T}}{dI/dV|_{0T} + dI/dV|_{2T}}$)[45] from site to site. The results demonstrate that the magnetic moment $M_S$ alters its orientation within the basal plane at the three different locations. To explicitly visualize the spatial distributions of the spin polarization, we present the differential conductance mapping at *B* = 0 and 2 T in Fig. 3e (the sum conductance resembles spin-averaged imaging, as shown in supplementary Fig. S8). The line profile of *P* in Fig. 3f indicates that the spin-up density is concentrated at the DH site, exhibiting a relatively sharp peak, while the other regions are concentrated with spin-down electrons. The maximal negative *P* appears at the BH site, while the

center site between BH and DH in the CDW supercell possesses a small negative $P$, characterizing the √19 × √19 CDW phase to be ferrimagnetic. Consequently, the magnetization orientation is modulated by the alternations of the structural chirality, while the alternating spin polarizations are synchronized with the alterchirality of the structural or charge order.

To further elucidate the magnetic properties of the √19 × √19 CDW phase, we investigate the effect of the on-site Coulomb repulsion $U$ within DFT calculations. We find, again, that the ferrimagnetic ground state emerges only when $U$ is large enough, while the √19 × √19 CDW phase remains nonmagnetic when $U$ is lower than 2.5 eV. In Fig. 4a and 4b, we plot the projected band structure and density of states (DOS) of the CDW phase without and with spin polarization, respectively, calculated with $U$ = 4 eV. Compared with the nonmagnetic case, the energy gain resulting from magnetization and time-reversal symmetry breaking amounts to 180.12 meV per supercell. Consequently, we observe spin splitting of electronic states near the Fermi level, as illustrated in Fig. 4b. In real space, this spin splitting also leads to a reduction in charge density of the itinerant electrons, corresponding to the states in the energy window of [$E_F$ - 0.08 eV, $E_F$] (supplementary Fig. S9). Additionally, from Fig. 4a and 4b, we note a strong hybridization between the Nb-4$d$ and Te-5$p$ orbitals, as well as the metallic nature characterized by the bands crossing the Fermi level. The resultant fractional magnetization is approximately 2.55 $\mu_B$ per supercell. These findings indicate that the ferrimagnetism exhibits an itinerant nature, at least partially.

Next, we examine the likely different contributions to the ferrimagnetism as well as its alternating nature in the $\sqrt{19} \times \sqrt{19}$ CDW phase. We start with eight initial candidate spin configurations, as shown in supplementary Fig. S10, and fully minimize their energies. We find that the two most stable magnetic configurations are degenerate, and exhibit completely opposite spin alignments (supplementary Fig. S10i and S10j). As an example of one of these two configurations, the atomic-scale distributions of the magnetic moments at the center site resemble well those of the SoD cluster in the $\sqrt{13} \times \sqrt{13}$ CDW phase, where the magnetization orientations of the surrounding Nb atoms are anti-parallel to that of the central one (see Fig. 4c). In contrast, in the BH or DH region, the spin directions of the central three Nb atoms keep mutually parallel (see Fig. 4c), consistent with the expectation of the $P_3$ space group of the $\sqrt{19} \times \sqrt{19}$ superstructure. In addition, the spin configurations between the BH and DH sites are always anti-parallel to each other, as displayed in Fig. 4c and 4d. Overall, these findings agree well with the *dI/dV* spectra and spin-polarized STM images shown in Fig. 3. Furthermore, the spin ordering is of a hybrid ferrimagnetic nature, primarily contributed by the itinerant electrons demonstrated above and localized *d*-orbitals here.

In conclusion, we have unraveled the correlation-driven quantum orders intertwined with lattice, charge, and magnetization in NbTe$_2$ at the monolayer limit. Most strikingly, a long-range alterchirality order, which is so far absent in crystalline monolayers, not only has been realized in the lattice and charge degrees of freedom, but is also synchronized with the magnetization. The alternating ferrimagnetism has been further shown to stem from hybrid contributions of the itinerant electrons and localized *d*- orbitals of the Nb atoms. These findings manifest a rare opportunity to explore the rich physics of chirality and alterchirality in correlated electron systems that exhibit mutually coupled multi-degrees of freedom, with important application potentials in sight.

## Methods

**Growth of the ML NbTe₂ thin films.** A NbTe$_2$ monolayer was grown on an epitaxial bilayer graphene (BLG)-terminated 6H-SiC(0001) substrate in an ultrahigh vacuum (UHV) molecular beam epitaxy chamber with a base pressure of ~2×10$^{-10}$ Torr. High-purity Nb (99.5%) and Te (99.999%) were co-evaporated on the BLG substrate[46] from an electron-beam evaporator and a standard Knudsen cell, respectively. The flux ratio was Nb:Te = 1:20, and the substrate was held at 350 °C during the film deposition. After growth, the ML NbTe$_2$ films were annealed at 400 °C for 2 h to improve the sample quality.

**STM/S measurements.** All STM/S measurements were performed on a commercial Unisoku 1300 LT-STM system at 4.8 K under UHV conditions. For spin-averaged STM/S measurements, an electrochemically etched tungsten tip was used and calibrated on a clean Cu(111) surface before the measurements. For spin-polarized STM/S measurements, 40-nm thick Cr layers were deposited on a clean W tip, followed by annealing at 500 K for 10 min. Prior to the Cr deposition, the W tips were flashed to ~2000 K to remove the oxides on the tips. The spin-polarized properties of the Cr-coated tips were examined on standard magnetic samples of Co islands on Cu(111) (see Supplementary S11). The *dI/dV* spectra and conductance maps at a constant tip-sample distance were taken by using a standard lock-in technique with a bias modulation frequency of 973 Hz.

**Density functional theory (DFT) calculations.** All first-principles calculations were performed within DFT implemented in the Vienna *ab initio* simulation package (VASP)[47] and Quantum Espresso,[48] using the generalized gradient approximation (GGA) of Perdew, Burke and Ernzerhof (PBE)[49] as the exchange-correlation functional. For VASP calculations, a 3×3×1 Monkhorst-Pack $k$-mesh[50] was used to sample the first Brillouin zone. The core electrons were treated fully relativistically by the projector augmented wave method[51] while the valence electrons were processed in the scalar relativistic approximation, with a plane-wave cutoff of 500 eV. All the atoms were allowed to fully relax during structural optimization until all the forces on each atom were less than 0.01 eV/Å. The DFT+$U$ formalism[52] was implemented within VASP calculations, where the $U$ value given to the Nb atom was 4.0 eV. To capture the lattice-driven reconstruction, we chose a Gaussian smearing of 0.02 eV. The phonon spectrum of 1$T$-NbTe$_2$ was obtained by adopting density functional perturbation theory[53] with a Gaussian smearing of 0.002 Ry on a 12×12×1 $q$-mesh, implemented within Quantum Espresso calculations.

## Acknowledgements

This work was supported by the National Key R&D Program of China (Grant Nos. 2018FYA0305800 and 2018YFA0703700), the National Natural Science Foundation of China (Grant Nos. 11974012, 12134011, 11974323, 12004368, and 12374458), the Strategic Priority Research Program of Chinese Academy of Sciences (Grant Nos.

XDB30000000 and XDB0510000), the Innovation Program for Quantum Science and Technology (Grant No. 2021ZD0302800), and the China Postdoctoral Science Foundation (Grant No. 2020M671859).

## Author Contributions

P.C., and C.D.Z. coordinated the research project; Y.S.B. prepared samples; Y.S.B. and J.H.D. performed STM experiments and analyzed data with support from T.J., C.Z., Z.M.P., D.H. and Z.B.C.; C.D.Z. and C-K.S. supervised the experimental measurements; G.H.C., H.M.F. and L.Q.L. carried out theoretical calculations under the supervision of P.C. and Z.Y.Z; C-K.S. helped with the discussion of the manuscript; Y.S.B., G.H.C., P.C., C.D.Z. and Z.Y.Z. wrote the manuscript with input from all other authors.

## Competing interests

The authors declare no competing interests.

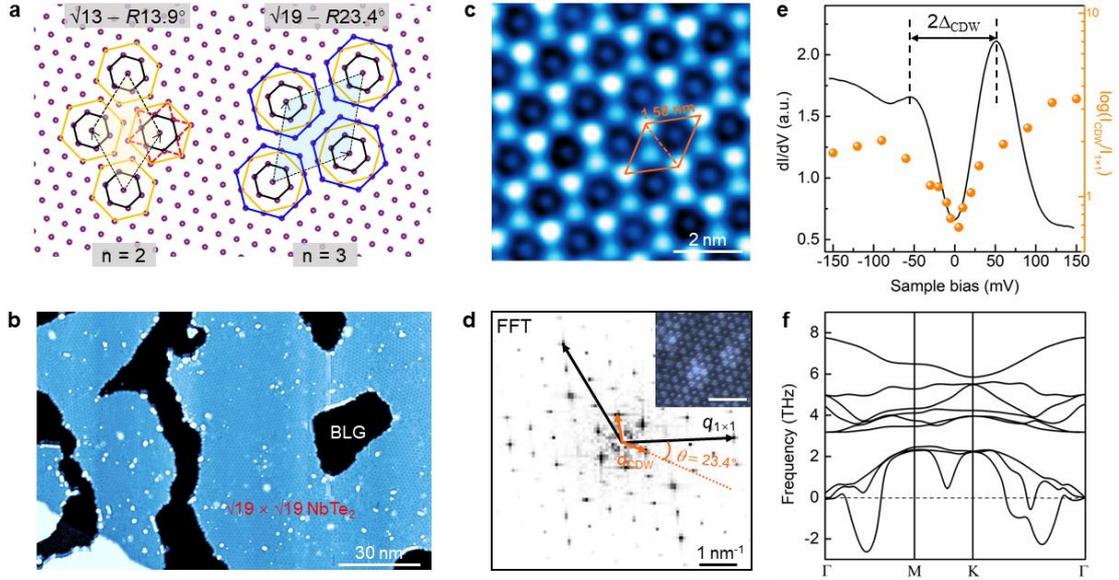

**Figure 1| Characterization of the √19 × √19 CDW phase in ML 1$T$-NbTe$_2$. a,** Schematic diagram of the 6$n$ + 1 stars-of-David systems and the relative rotation angles between the atomic lattices and CDWs, with the purple spots representing the transition metal atoms. **b,** Large-scale STM morphology of the ML 1$T$-NbTe$_2$, ~~where the sample surface is fully covered with the √19 × √19 superstructure,~~ the sample surface is dominated by the √19 × √19 superlattice. **c,** Typical STM topography of the √19 × √19 superstructure under negative bias, with the orange rhombus corresponding to the CDW supercell. **d,** FFT of the atomic resolution STM image (inset of d) of the √19 × √19 superstructure, with the black and orange arrows marking the Bragg and CDW vectors, respectively. The scale bar in the inset is 4 nm. **e,** Typical $dI/dV$ spectrum of the √19 × √19 CDW (black line), superimposed with the relative CDW intensity (orange dots), both as functions of the energy, where 0 mV refers to the Fermi level. **f,** Calculated phonon spectrum of a pristine NbTe$_2$ monolayer along high symmetry lines. Scanning parameters: in **b,c,** $V_b$ = -1 V, $I_t$ = 10 pA; in the inset of **d,** $V_b$ = -1 V, $I_t$ = 500 pA; in **e,** $V_b$ = -300 mV, $I_t$ = 200 pA before turning off the feedback loop, and $V_{mod}$ = 3 mV.

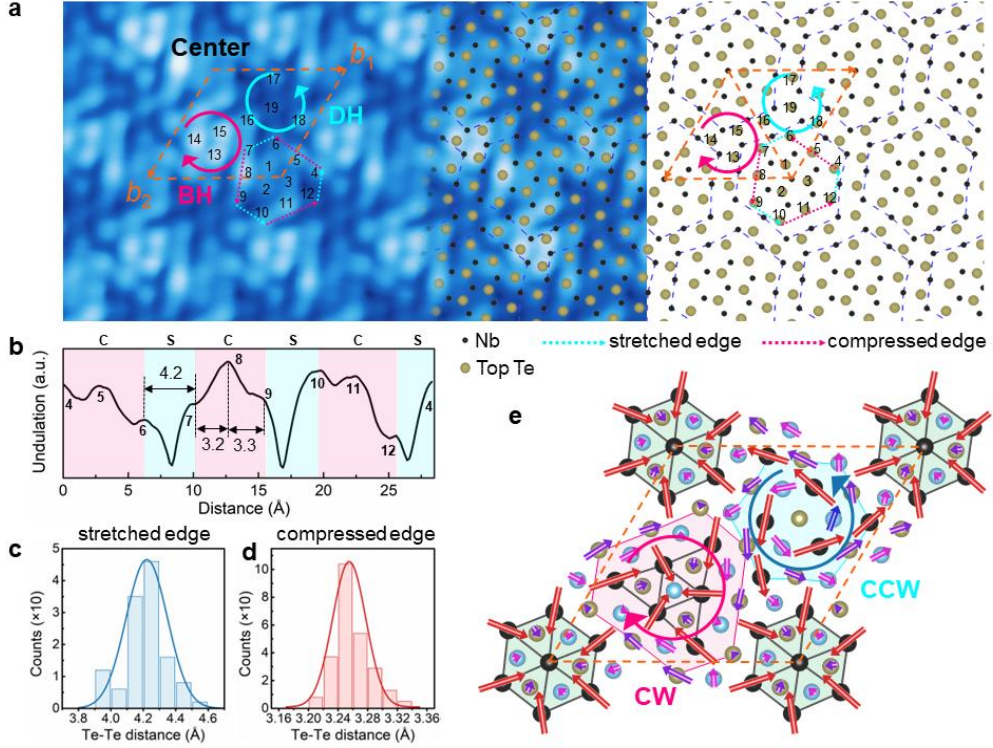

**Figure 2| Alterchirality in the √19 × √19 CDW phase. a,** Atomically resolved STM image taken on the √19 × √19 CDW phase in a NbTe$_2$ monolayer, superimposed with the DFT-calculated atomic structure. In the latter, the topmost Te atoms and middle Nb atoms are labeled by the brown and black dots, while the bottom Te atoms are not shown for better clarity. Each 19-Nb cluster is highlighted by the blue dashed lines, and the top 19 Te atoms within a CDW supercell are indexed as #1 to #19. Scanning parameters: $V_b$ = +10 mV, $I_t$ = 3 nA. **b,** Line profile measured along the three compressed (pink) and stretched (cyan) edges in **a**. The capital letters C and S on the top correspond to the compressed and stretched edges, respectively. **c,d,** Histograms of the inter-atomic distances of the stretched and compressed edges. The solid lines are fits to Gaussian normal distributions. **e,** Detailed structural distortions of the √19 × √19 supercell from the DFT calculations, forming two sites with opposite chirality (e.g., clockwise within the BH site and counterclockwise within the DH site). The Nb, top Te, and bottom Te atoms are represented by the black, brown, and cyan balls, respectively. The arrows represent the orientations of the atomic displacements; the length of each arrow indicates the magnitude of the displacement amplified by 3.8.

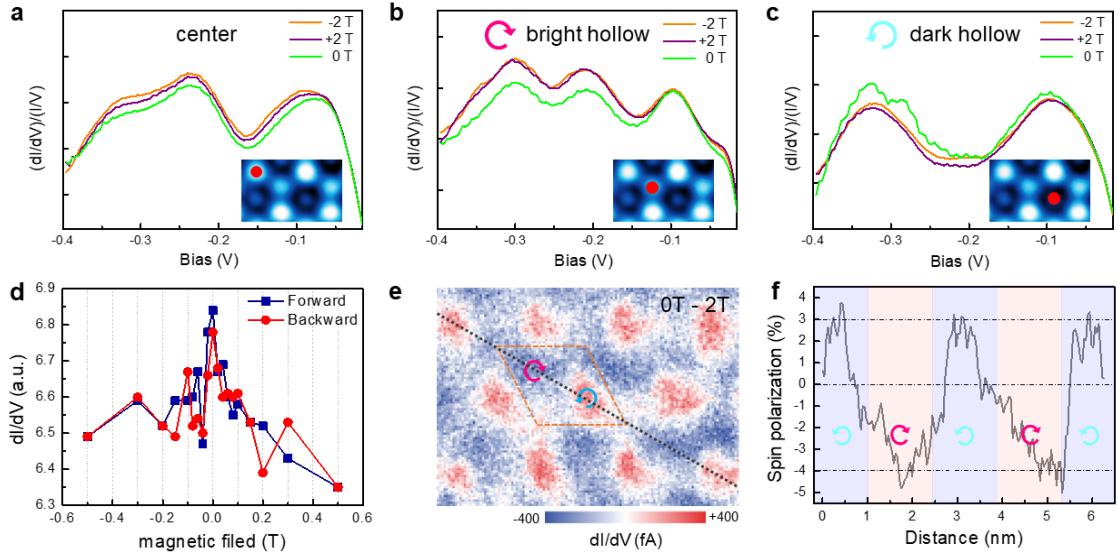

**Figure 3| Spin-polarized tunneling spectra of the √19 × √19 CDW phase in ML NbTe$_2$. a-c,** Normalized spin-polarized *dI/dV* signals of the center, BH, and DH sites, each obtained under perpendicular external fields of ±2 and 0 T. **d,** *dI/dV* signal as a function of the magnetic field, as extracted from the DH site at $V_b$ = -0.35 V. The out-of-plane field is swept forward (blue) and backward (red). **e,** Differential conductance mapping of the same area at 0 and 2 T. **f,** Line profile of spin polarization *P* along the black dashed line in **e**, displaying spatial alternations in the magnetization. Tunneling parameters: **a-c,** $V_b$ = -0.4 V, $I_t$ = 100 pA before turning off the feedback loop, and $V_{mod}$ = 5 mV.

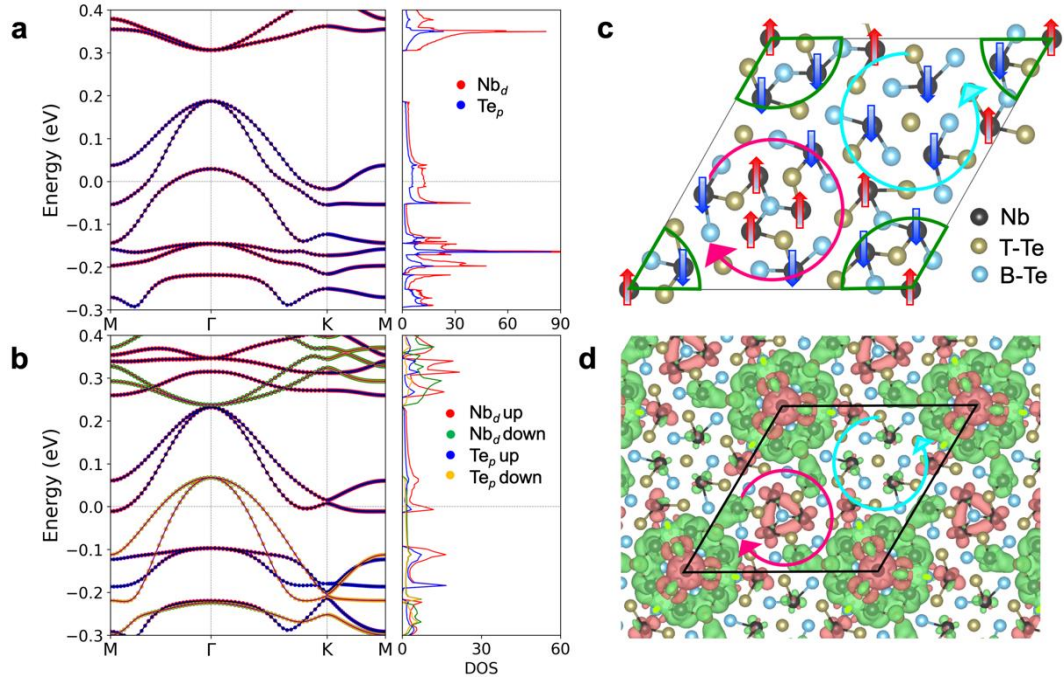

**Figure 4| First-principles calculated electronic and spin structures of the √19 × √19 superstructure. a,b,** Orbital-decomposed band structures and densities of states with and without spin polarization, respectively. **c,** Spin configuration of one of the two degenerate stable structures within a supercell. The blue, red, and green arcs indicate the DH, BH, and center sites, respectively. The blue and red arrows denote spin down and up components at the Nb atoms, respectively. **d,** Spin-resolved charge density distribution, with the green and red shadows representing the spin down and up density, respectively.